\documentstyle [12pt]{article}
\title{Anomalous scaling in random shell models for passive~scalars}
 \author{A. Wirth$^{1}$ \& L. Biferale$^{2}$}

\begin{document}

\begin{center}
\maketitle
\centerline{${^1}$ CNRS, Observatoire de Nice, B.P. 229, 06304 
Nice Cedex 4, France.}
\centerline{${^2}$ Universit\'a di Roma ``Tor Vergata'',  Dip. di Fisica}
\centerline{Via della Ricerca Scientifica 1,
00133 Roma, Italy. }
\date{ }
\medskip
\medskip
\end{center}
\vspace{0.2truecm}
\noindent

\begin{abstract}
A  shell-model version of Kraichnan's (1994 {\it
Phys. Rev. Lett.  \bf 72}, 1016) passive scalar problem is introduced
which is inspired from the model of Jensen, Paladin and Vulpiani
(1992 {\it Phys. Rev. A\bf 45}, 7214).
As in the original problem, the prescribed random velocity field is
Gaussian, delta-correlated in time and has a power-law spectrum 
$\propto k_m^{-\xi}$, where $k_m$ is the wavenumber. 
Deterministic differential equations for second and fourth-order
moments are obtained and then solved numerically. 
The second-order structure function of
the passive scalar has normal scaling, while the fourth-order
structure function has anomalous scaling. For $\xi = 2/3$ the anomalous scaling
exponents $\zeta_p$ are determined for structure
functions up to $p=16$ by Monte Carlo simulations of  the random shell model,
using a stochastic differential equation scheme, validated 
by comparison with the results obtained for the second and
fourth-order structure functions.
\end{abstract}
{\bf \hfill ROM2F/96/29}\\

\medskip
 
\section{Introduction}
\label{s:introduction}

A most striking property of a passive scalar quantity advected by a
fully developed turbulent flow is its spatio-temporal intermittent
behavior (see Ref.~\cite{SREENE} for a recent review).  The time
evolution of the scalar field $\theta$ is described by the partial
differential equation\,:
\begin{equation}
\partial_t \theta + ({\bf v}\cdot  \nabla) \theta = \kappa \Delta \theta + f,
\label{eq:pde}
\end{equation}
where ${\bf v}$ is the advecting velocity field, $\kappa$ is the
molecular diffusivity and $f$ is an external forcing. 

Intermittency in passive scalars advected by highly chaotic flows is
thought to be connected to the existence of self-similar processes
transferring fluctuations from large to small scales.

Self-similarity is observed in the power-law behavior of structure
functions, $\tilde S_p(r)$, in the inertial range, that is moments of
$\theta$-increments at scales where neither external forcing nor
molecular damping are acting\,:
\begin{equation}
\tilde{S}_p(r)  \equiv  \langle|\theta(x)-\theta(x+r)|^p\rangle 
\sim  r^{\zeta_p}.
\label{eq:sf}
\end{equation}
The set of scaling exponents, $\zeta_p$, fully characterizes
intermittency.  In particular, deviations from the dimensional
(linear) behavior $\zeta_{2p}=p \, \zeta_2$ are evidence of a
nontrivial scalar transfer among scales, in analogy to the energy
intermittent cascade in turbulent flows (see Ref.~\cite{UF} for a recent
review on this subject).

Another important issue  is  connected to the questioned universal
character of scalar fluctuations in the inertial range. Universality
should be the consequence of  local, or quasi-local, interactions
among scales  and it should be reflected in a strong robustness
of the scaling exponents, $\zeta_p$, against variation of
forcing and/or dissipation mechanisms. This idea dates back to the 
phenomenological work of Obukhov and
Corrsin \cite{OC}. 

Statistics
of the passive scalar must be strongly related to the properties
of the advecting velocity fields; therefore,  the only 
realistic case would be to study the ``full problem'' given by 
(\ref{eq:pde}) together with  the corresponding Navier--Stokes equations,
in the limit of high Reynolds number,  which describes
the time evolution of  a fully developed turbulent  velocity field.

Due to the lack of knowledge about the statistical properties
of a fully developed turbulent flow the ``full problem'' 
cannot yet be attacked. Nevertheless, equation (\ref{eq:pde})
with a prescribed velocity field of suitable statistical properties,
could be a good ``training ground'' for understanding general
mechanisms (if any) underlying the presence of
anomalous scaling in fluid-dynamics.

Recently, much attention has been paid in this direction by studying
the intermittency properties of a passive scalar advected by a
velocity field which is Gaussian, white-in-time and which has
self-similar spatial correlations.  In
Refs.~\cite{K94,GK,KYC,CFL,CFL2,E,GKB} the two-point velocity field
correlation function was set to\,: $\langle v_i(x,t)
v_j(x',t')\rangle=\delta(t-t')D_{ij}(x-x')$, with $D(x) =
D(0)-\hat{D}(x)$. Here,  $\hat{D}$ is the $d$-dimensional velocity-field
structure function\,:
\begin{equation}
\hat{D}_{ij}(x) = D_0 |x|^{\xi} \left[ (d-1+\xi)\delta_{ij}
 - \xi x_ix_j |x|^{-2} \right],
\label{eq:corre}
\end{equation}
where the scaling exponent $\xi$ of the second order velocity
structure function, $\xi$, with $0 <\xi < 2$, is a free parameter.
Higher order velocity-field correlation functions are fixed by
the Gaussian assumption.

Such a choice is far from being
realistic because of the assumed  fast-time-decorelation and Gaussianity.
Therefore, no quantitative agreement with the intermittent properties
of the ``full problem'' must be expected.  Nevertheless, in
Refs.~\cite{K94,GK,KYC,CFL,GKB,P}
many interesting  analytical and phenomenological results for this toy-model 
have been obtained,
giving for the first time the impression that some new insight into the
``intermittency problem'' has been gained (see Ref.~\cite{massimo} for related
results in  a passive vector case). 

The big advantage of this toy-model is that, due to the
delta-correlation in time, all equations for equal-time $n$-points
correlations are formally closed, that is, the $n$-th-order
correlation depends only on lower order correlations. This fact was
stressed by Kraichnan \cite{K94,K68}, who gave for the first time the
closed expression for $\tilde{S}_2(r)$. In Ref.~\cite{K94} a
theory for all structure functions was proposed and an explicit
formula for $\zeta_p$ given.  The main physical outcome was that
all structure functions of order greater than two have intermittency
corrections and that intermittency should be connected to some non
trivial matching between advective and diffusive properties of the
model.  Indeed, for structure functions of order greater than two it
is no more possible to clearly separate the contributions from
diffusion and advection terms in the generation of anomalous scaling.

In Refs.~\cite{GK,CFL,GKB} it has been shown that intermittency of the
scalar structure functions is connected to the properties of the
null-space of the linear operators describing the inertial-scale
evolution of multipoint passive scalar moments. Moreover, perturbative
expressions of intermittency corrections, as a function of the
parameter $\xi$ and as a function of the inverse of the dimensionality
of the system were given \cite{GK,CFL,GKB}. In both cases
careful analysis of matching conditions at infrared (IR) and
ultraviolet (UV) scales must be taken into account.  Universality in
the scaling exponents is still preserved, while constants in front of
the power laws acquire dependencies on the forcing and on the passive
scalar dissipation. How far all these results can be freely extended
to the ``full problem'' is difficult to say. Certainly, the scaling
exponents have a strong quantitative dependence on the statistical
properties of the prescribed velocity field.  Influence of a
short-but-non-zero correlation time was studied in Ref.~\cite{CFL2}
and found important, while from a phenomenological point of view it is
difficult to say if the perturbation expansion of Ref.~\cite{GK} can
be extended up to the value $\xi=2/3$ which mimics the spatial energy
spectrum of true turbulence. (Let us notice that, even for $\xi=2/3$,
the assumed delta-correlation in time for the velocity makes the 
classical Obukhov-Corrsin theory inapplicable to the passive scalar
two-point correlation function \cite{K68}.)

An important drawback of the Gaussian delta-correlated toy-model is
that numerical simulations seem even more difficult than a direct
numerical simulation of the ``full problem''.  This is because
(\ref{eq:pde}) is a stochastic differential equations (SDE) the
solution of which requires huge computer resources, because of the
massive use of random number generators and also because of the
paramount difficulties in the implementation of higher than first
order discretization schemes \cite{KP}. The state of the art of SDE
numerical simulations is hardly compatible with the high spatial
resolution needed to study scaling properties of the solution of
(\ref{eq:pde}) with a white-in-time velocity field (see however 
Ref.~\cite{KYC} for a first  attempt at overcoming these
difficulties, albeit at the prize of introducing some anisotropy into
the velocity field).
 
In this spirit, we have tried to work with the simplest toy-model
which has some connection to the physics of the ``full problem'',
while at the same time aiming for something more tractable
analytically and/or numerically.  We have thus investigated the
intermittency properties of a shell-model for a passive scalar
advected by a prescribed stochastic velocity field. Shell models (see
Ref.~\cite{kada} for a pedagogical introduction) have been already
successful in helping to understand many issues connected to fully
developed turbulence
\cite{G,YO,PVJ2,BBP,KLWB,BBKT}.  

The problem of defining a shell model for the advection of a passive
scalar by a deterministic and chaotic velocity field has been already
investigated in Ref.~\cite{PVJ}. Here, we are going to present our
numerical and analytical results obtained in a weakly modified
version of the shell model defined in Ref.~\cite{PVJ}, but using a
delta-correlated surrogate for the advecting velocity field.  The
main drawback of our model is that, following shell-model philosophy,
it retains only nearest and next-nearest neighbor interactions in
the shell index.  Having only semi-local interaction in the Fourier
space could be a problem for describing the physics of passive
fluctuations for those types of velocity-field-correlations which
induce strong non-local effects. For example, simple dimensional
considerations tell us that the limit of very small $\xi$ should be
dominated by strong non-local effects (which leads to an ultraviolet
divergence of the eddy-diffusivity in the limit $\xi
\rightarrow 0$ \cite{GK}). Therefore we do not expect that our shell model
could properly mimic the physics of the PDE (\ref{eq:pde}) in the
range of parameters choice $0<\xi \ll 1$. Yet, we are quite confident
that our model captures very well the passive scalar physics for $\xi
= O(1)$, that is when dominant interactions become more local in
Fourier space, as is the case for true turbulence. The major advantage
of using a shell-model is that, now, reliable numerical simulations
become feasible, allowing nonperturbative determinations of scaling
exponents up to high orders (in this paper we present results up to
order $16$). Furthermore, having reliable numerical simulations should
open the possibility of relaxing some of the constraints put on the
statistical properties of the velocity field, thereby allowing
investigation of increasingly realistic problems.

We found that our shell model shares many properties with the original
SDE (\ref{eq:pde}).  Among them, the most importants are: (i) the
second order structure function has normal scaling, (ii) all structure
functions of order larger than two have anomalous corrections, (iii)
there is a remarkable coupling between both UV and IR scales with
inertial terms in the closed equation satisfied by fourth-order shell
correlations.  This result suggests that the anomalous scaling is due
to nontrivial contributions of integral and diffusive scales on the
inertial-range properties.  Moreover, we find that anomalous behavior
tends to vanish when approaching the laminar regime, $\xi=2$, for the
advecting velocity field.

Even though we have not obtained an analytical expression for the
anomalous scaling exponents, we are confident that many useful insights can
be obtained by further investigation of the present model, or even simpler
one.  For example, it should be possible to define some
exactly solvable models, where the shell-velocity correlations is
chosen such as to give exactly solvable (but nontrivial) linear operators
describing the structure functions inertial properties

The paper is organized as follows\,: in Section~\ref{s:goveq} we
introduce the random passive scalar shell model and we discuss some
problems connected with its stochastic differential equation (SDE)
formulation.  In Section~\ref{:goveqen} we explicitly derive the
closed deterministic equations for second and fourth-order
moments, which are numerically solved in Section
\ref{s:Deterministic} to obtain the scaling exponents of second and
fourth-order structure functions. Monte Carlo simulations of the SDE
are presented in Section~\ref{s:Stochastic}.  Concluding remarks and
suggestions for further work are given in Section~\ref{conclusion}.


\section{The model and its stochastic differential equation formulation}
\label{s:goveq}
We recall the main features of the passive scalar shell model of 
Ref.~\cite{PVJ}.
The model is defined in terms of a shell-discretization
of the Fourier space in a set of wavenumber defined on a geometric
progression $k_n = k_0 2^n$. Passive increments at scale $r_n=k_n^{-1}$
are described by a complex variable $\theta_n(t) $. The
time evolution is obtained according to the following criteria\,: (i)
the linear term is a purely diffusive term given by $-\kappa k_n^2 \theta_n$;
(ii) the advection term is a combination of the form
 $k_n \theta_{n'} u_{n''}$; (iii) interacting shells are restricted to
nearest and next-nearest neighbors of $n$; (iv) in the absence of forcing
and damping the model conserves the volume in the phase-space and 
the passive-energy $ E = \sum_n |\theta_n|^2$. Properties (i), (ii) and 
and (iv) are valid also for the original equation (\ref{eq:pde})
in the Fourier space, while property (iii) is an assumption
of locality of interactions among modes, which is rather well
founded as long as the power law spectra of the passive scalar $E(k) \sim
k^{-\alpha}$ has $1 < \alpha < 3$. 

Our passive scalar shell model, inspired by Jensen, Paladin and Vulpiani's
model \cite{PVJ}, 
is defined by the following equations ($m=1,2,\ldots$)
\begin{eqnarray}
[\frac{d}{dt} + \kappa k_m^2] \theta_m (t) &=& 
   i [a_{m} 
(\theta_{m+1}^* (t)  u_{m-1}^*(t)
-\theta_{m-1}^* (t) u_{m+1}^* (t)) \nonumber \\
& + & b_{m}
(\theta_{m-1}^* (t)   u_{m-2}^* (t)
+\theta_{m-2}^* (t)  u_{m-1}(t)) \nonumber \\
 &+  &c_{m}
(\theta_{m+2}^* (t) u_{m+1}(t)
+\theta_{m+1}^* (t)  u_{m+2}^* (t))]
 + \delta_{m1} f(t),
\label{shellmodel}
\end{eqnarray}
where the star denotes complex conjugation and
\begin{equation}
k_{m} = 2^m,  \;\;
 a_{m} =  \frac{k_{m}}{2},\;\; 
 b_{m} = -\frac{k_{m-1}}{2},\;\; 
 c_{m} =  \frac{k_{m+1}}{2},
\label{abc}
\end{equation}
with $u_{-1}=u_0=\theta_{-1}=\theta_0=0$ as boundary conditions. 
The forcing term $\delta_{1m} f(t)$ 
acts only on the first shell. Note that (\ref{abc}) is one of
possible choices for the parameters $a_{m}$, $b_{m}$ and $c_{m}$ ensuring the
conservation of passive scalar energy $\sum_m \theta_m \theta_m^*$
and phase-space volume.
In numerical implementations, the model is 
truncated to a finite number 
of shells $N$ (here $N=19$) with the additional boundary conditions 
$u_{N+1}=u_{N+2}=\theta_{N+1}=\theta_{N+2}=0$.

Our model differs from that of Ref.~\cite{PVJ} by the absence of 
complex conjugation on two of the $u$ factors in the rhs of 
(\ref{shellmodel}). The reason for this change will become clear below.
Furthermore, in Ref.~\cite{PVJ} the passive scalar model 
(\ref{shellmodel}) was coupled to the GOY shell model \cite{G,YO,PVJ2}
for the nonlinear dynamics of the velocity variables.
The GOY model displays multifractal behavior for the $u_n$'s
and an interesting intermittent behavior is also found for the passive scalar.

Our goal,as in Kraichnan's work \cite{K94}, is to use a {\it
nonintermittent\/} velocity field and then to find if the passive scalar
is nevertheless intermittent.  For this, we assume that the
velocity variables $u_m (t)$ and the forcing term $f(t)$ are
independent complex Gaussian and white-in-time, that is,
delta-correlated.  Furthermore, as in Ref.~\cite{K68}, we make a
scaling assumption for the spectrum of the $u_m (t)$'s, namely
\begin{eqnarray}
\langle u_m (t) u_m^* (t') \rangle = \delta(t-t') D_m,
\end{eqnarray}
where
\begin{eqnarray}
D_m= k_m^{-\xi}
\end{eqnarray}
will be called the velocity spectrum. For $f(t)$ we assume 
$\langle f(t) f(t') \rangle = \delta(t-t')$.

As long as the velocity variables $u_m (t)$ have a finite correlation time
and, hence, smooth sample paths, there is no particular difficulty in 
giving a meaning to the set (\ref{shellmodel}) of random ordinary 
differential equations
(ODE's). A well-known difficulty arises with ODE's having white-in-time 
coefficients\,: the mathematical meaning of the equation is 
ambiguous \cite{KP}.
The physicist's and thus our view-point is to define the solution of such
``stochastic differential equations'' (SDE's) to be the limit, as the
the correlation time tends to zero, of the solution of a random ODE with
nonwhite (colored) coefficients. 
In the mathematical literature on stochastic differential equations 
this is called the Stratonovich version of stochastic calculus \cite{KP}.
When numerically solving a SDE such as (\ref{shellmodel}) one cannot 
use standard numerical methods which assume {\it smooth\/} coefficients. 
It would also be highly impractical to use a smooth approximation of 
the coefficients with a small correlation time, since this requires 
time steps much smaller than the correlation time.
Fortunately, there is an alternative formulation of the SDE, 
the Ito version, which overcomes this difficulty.
The solutions of SDE's are Markov diffusion process which can be
characterized by their diffusion and drift coefficients 
(also called Fokker-Planck coefficients). 
The so-called Ito equation \cite{KP} encodes this information in a way which
leads naturally to efficient numerical schemes.

Let us briefly explain how this Ito equation is obtained.
For this it is better to work with an abstract form of the 
starting equation (with the forcing term omitted)\,:
\begin{eqnarray}
\frac{d \theta}{dt} = {\bf M_0}\, \theta +\frac{1}{\epsilon} 
{\bf M_1} (\frac{t}{\epsilon^2})\, \theta.
\label{www}
\end{eqnarray}
Here, ${\bf M_0}$ is a deterministic operator and ${\bf M_1}$ a random Gaussian
operator with finite correlation time. The $\epsilon$-scaling
is chosen in such away that, for $\epsilon \rightarrow 0$, the
${\bf M_1}$ operator becomes delta-correlated in time.
Just as  ordinary white noise may be written as the derivative, in the sense 
of distributions, of the Brownian motion process, we can write
\begin{eqnarray}
\lim_{\epsilon \rightarrow 0} \,\frac{1}{\epsilon} {\bf M_1}
(\frac{t}{\epsilon^2})= 
\frac{{\bf dW}}{dt},
\label{mmw}
\end{eqnarray}
where ${\bf W(t)}$ is an operator-valued Brownian motion.
The Ito SDE associated to (\ref{www}) is 
\begin{eqnarray}
d \theta = ({\bf M_0} +{\cal D}) \theta+ {\bf dW}\theta,
\label{tt}
\end{eqnarray}
where
\begin{eqnarray}
{\cal D}=\int_0^{\infty} \langle {\bf M_1(s)} {\bf M_1(0)} \rangle ds
\label{DD}
\end{eqnarray}
is called the drift operator.
The presence of the drift term in (\ref{tt}) is easily understood\,:
a Neumann expansion of $\theta(t+dt)-\theta(t)$ in powers of
${\bf M}_1$, starting  from (\ref{www}), gives a second-order term, which is 
$O(dt^2)$ for fixed $\epsilon$ but becomes ${\cal D}\theta dt$ when the limit 
$\epsilon \rightarrow 0$ is taken.
The drift term may also be obtained by using the Ito formula \cite{KP} or
Gaussian integration by parts (see, e.g., Ref.~\cite{UF}, Section 4.1).
The term ${\bf dW}\theta$ is called the diffusion term.
Since Brownian motion has independent increments over nonoverlapping intervals,
the operator ${\bf dW}$, which is $O(\sqrt{dt})$ and has zero mean, is
independent of the past and, thus, of $\theta$.
It follows that the diffusion term has vanishing mean.
Hence, the mean of $\theta$ satisfies
\begin{eqnarray}
\frac{d \langle \theta \rangle}{dt} = ({\bf M_0} +{\cal D})\,
\langle \theta \rangle.
\label{DDD}
\end{eqnarray}

The Ito SDE associated to (\ref{shellmodel}) reads\,:
\begin{eqnarray}
d \theta_m (t)& = & 
((a_{m}b_{m+1}+b_{m}c_{m-2}) D_{m-1} +(-a_{m}c_{m-1}
+b_{m+2}c_{m}) 
 D_{m+1}
\nonumber \\
& +  & a_{m-1}b_{m} D_{m-2}-a_{m+1}c_{m} D_{m+2}- 
\kappa k_m^2 )
\theta_m (t) \, dt
\nonumber \\
& +  &i [a_{m} 
(\theta_{m+1}^* (t) \alpha_{m-1} dW_{m-1}^* (t)
-\theta_{m-1}^* (t) \alpha_{m+1} dW_{m+1}^* (t))
\nonumber \\
& +  &b_{m}
(\theta_{m-1}^* (t)   \alpha_{m-2} dW_{m-2}^* (t)
+\theta_{m-2}^* (t)  \alpha_{m-1} dW_{m-1} (t))
\nonumber \\
& +  &c_{m}
(\theta_{m+2}^* (t) \alpha_{m+1} dW_{m+1}(t)
+\theta_{m+1}^* (t)  \alpha_{m+2} dW_{m+2}^* (t))]
\nonumber \\
& +  & \delta_{m,1} dW_{f}(t).
\label{shelleqito}
\end{eqnarray}
Here, $\alpha_m=k_m^{\xi/2}$, $D_m=\alpha_m^2/2$, the $W_m(t)$'s and
$W_f(t)$ are independent identically distributed complex-valued
Brownian motion functions, normalized in such a way that $\langle |
W_m (t) |^2 \rangle = \langle | W_f (t) |^2 \rangle =t$.  It is
noteworthy that the drift term in (\ref{shelleqito}), the top two
lines on the rhs, which involves the sum of the energies in the
neighboring and next-neighboring shells, may be viewed as an
eddy-diffusivity term. For Kraichnan's original
equation (\ref{eq:pde}), the Ito equation approach has a
scale-independent eddy diffusivity \cite{FW}.  Here, it is
proportional to $ k_n^{-\xi}$. The difference  stems
from the absence, in the shell model of direct interactions between
widely separated scales.  We also observe that, if we had used the
original model of Ref.~\cite{PVJ} with complex conjugates on all the $u$
factors, the drift operator would involve nondiagonal elements
coupling different shells, a situation we avoided.


\section{Equations for the second and 
fourth-order moments}
\label{:goveqen}

In this and the following sections we are interested in the scaling behavior
of the $p$-th-order structure functions\,:
\begin{eqnarray}
\langle (\theta_m \theta_m^*)^{p/2} \rangle \propto k_m^{-\zeta_p},
\label{struct}
\end{eqnarray}
where $\zeta_p$ is called the scaling exponent of order $p$.
If $\zeta_{2p}=p \, \zeta_2$ the structure functions are said to have
a normal scaling.
If $\rho_{2p} =\zeta_{2p}-p\, \zeta_2 \neq 0$ 
the scaling of the structure function of order $2p$ is said to be anomalous.

It is well known that from a linear stochastic differential equation
with white-noise coefficients it is possible to obtain exact equations
for moments of arbitrary order \cite{K68,KYC}.  For example, from the
abstract equation (\ref{www}) one derives the closed equation for the
first-order moment (\ref{DDD}).  Higher order quantities such as
$\theta \otimes \theta$, $\theta \otimes \theta\otimes \theta \otimes
\theta$, \ldots satisfy also linear stochastic differential equations
with white-noise coefficients, from which closed equations can be
obtained for $\langle \theta \otimes \theta \rangle$, $\langle \theta
\otimes \theta\otimes \theta \otimes \theta \rangle$, etc. .  Such
equations may also be obtained by use of Ito calculus \cite{KP}.

In the shell-model context the moment equations become excessively cumbersome
beyond order four.
We have obtained the closed equations for
\begin{eqnarray}
E_{m} \equiv \langle \theta_m  \theta_m^* \rangle \quad {\rm and} \quad
P_{lm}=\langle \theta_l \theta_l^* \theta_m \theta_m^* \rangle.
\end{eqnarray}
The general structure of these equations is as follows
\begin{eqnarray}
\dot{ E}_l&=&(-2 \kappa k_m^2  \delta_{l,m}+ A_{l,m}) E_m + F_l.
\label{E}\\
\dot{P}_{lm}&=& [ -2 \kappa (k_l^2+k_m^2)\delta_{l,n}\delta_{m,j}+
 B_{lm,nj} ] P_{nj} +G_{lm}. 
\label{s4}
\end{eqnarray}
The equations are written in explicit form in the Appendix~B.
Here, we stress a few important properties.
The forcing $F_m$ is restricted only to the first shell $m=1$.
The matrix $A_{l,m}$ is symmetric band diagonal with bandwidth five and
has the following scaling law\,:
\begin{eqnarray}
A_{l+s,m+s} = k_s^{2-\xi} A_{l,m}.
\end{eqnarray}
Similarly,
\begin{eqnarray}
B_{l+s;m+s,n+s;j+s}= k_s^{2-\xi} B_{lm,nj}
\end{eqnarray}

Straightforward scaling arguments indicate that (\ref{E}) and (\ref{s4})
may possess steady-state solutions with {\it normal} scaling.
Such solutions have 
\begin{eqnarray}
E_m \propto k_m^{2-\xi} = k_m^{-\zeta_2}
\end{eqnarray}
and
\begin{eqnarray}
P_{mm} \propto k_m^{-\zeta_4}
\end{eqnarray}
with $\zeta_4=2 \zeta_2$.
In order to find what kind of scaling actually holds, we now resort 
to numerical solutions of the moment equations (\ref{E}) and (\ref{s4}).
Since these are {\it deterministic\/} equations they can be solved with 
high accuracy at relatively low cost. This is not the case of the 
Monte Carlo strategy of Section~5, which allows us however, to 
tackle structure functions of high orders.


\section{Calculation of second and fourth-order structure functions 
from the moment equations}
\label{s:Deterministic}


For the numerical solution of (\ref{E}) and (\ref{s4}) we used a $19$-shell
truncation with values of $\kappa$ chosen in such a way that the 
diffusive cutoff is well within the available range of shells.
The solutions were obtained by time-marching until the steady state
is reached. Such a steady state is necessarily stable.
The scaling parameter $\xi$ was varied in the range $0.2 \leq \xi \leq 2$.

We found that the second order structure function $E_m$ displays 
always normal scaling with
\begin{eqnarray}
\zeta_2=2-\xi
\end{eqnarray}
the same value as obtained by Kraichnan \cite{K68}.  To investigate
the nature of the scaling of the fourth-order structure function
$P_{mm}$, we plotted it against the second-order structure function,
following the, now standard, Extended Self Similarity (ESS) procedure
(see, Ref.~\cite{Benzi}). This gave us the anomalous part
$\rho_4=\zeta_4 - 2 \zeta_2$ of the scaling exponent, plotted in
Fig.~1 against $\xi$.  It is clearly seen that the scaling is
anomalous ($\rho_4 > 0$), the anomaly is a decreasing function of
$\xi$ and disappears as $\xi \rightarrow 2$, the ``laminar'' limit, as
in Ref.~\cite{K94}.

For small values of $\xi$, interactions become more and more nonlocal
and a higher number of shells would be needed to have a sufficiently
large inertial range. (The number of shells in the ``intermediate
inertial-dissipation range'', in which the local Reynolds number is
$O(1)$, grows like $1/\xi$.)  Even though our shell model has only
local interactions, purely kinematic effects introduce important
long-range diffusive corrections in the limit $\xi \rightarrow 0$.
This is why we restricted our calculations of $\rho_4$ for $\xi$
ranging from $0.2$ to $2.0$.

The fact that normal scaling is obtained for the second-order
structure function is not very surprising, this being exactly the same
situation as for Kraichnan's original problem.  It is easily checked
that the operator $A$ appearing in (\ref{E}) has equipartition
solutions ($E_m$ independent of $m$) in its null-space, both for the
full operator and its $N$-shell truncations.  Such equipartition
solutions have no associated passive scalar energy flux and cannot
bring about anomalous scaling.

Let us now consider the anomalous scaling for the fourth-order
structure function which are connected to the properties of the
operator $B \equiv B_{lm,nj}$.  The operator $B$ has again
equipartition solutions in its null-space, which cannot cause anomalous
scaling.  We have checked numerically that finite-shell truncations of
$B$ have no other eigenvectors in their null-space.  This, only
superficially contradicts the interpretation of anomalous scaling as
arising from the zero modes of the inertial operator describing the
evolution of moments \cite{GK,GKB}. In our case there are no
anomalous zero-modes of the operator $B$ but still we see numerically
a very clean intermittent behavior. Where does the observed anomaly
come from? One can imagine two interpretations. The first one rests on
the physical observation that the operator $B$ is naturally
long-range\,: it mixes inertial scales with cut-offs at both the UV
and IR ends. Therefore, its inverse involves a nontrivial mixture of
contributions from very different scales.  This mixture could be the
cause of anomalous scaling, thereby defeating naive (local) dimensional
analysis.  The second interpretation is to imagine that quasi-zero
modes (which would become true zero-modes in the limit of an infinite
inertial range) are already dominating the inverse of $B$. The two
interpretations are not in contradiction. Indeed, our truncated system,
being always influenced by UV and IR cut-offs, it naturally takes into
account boundary conditions and therefore never shows true zero-modes.
A similar scenario takes place in Ref.~\cite{GK}, where zero modes in
the infinite dimensional function space have to be matched with IR and
UV physical cut-offs.

Let us finally remark, that  the  analysis
of the relevant eigenvectors in the eigenspace of the operator $B$  
 is highly complicated due to the fact that we are looking at the cone
of positive functions, which is not a linear space.


\section{Monte Carlo simulations for structure functions}
\label{s:Stochastic}


As we noted, the moment-equation strategy becomes inpractical for
determining structure functions beyond the fourth order.  We therefore
resort to Monte Carlo simulations of the stochastic shell model in its
Ito version (\ref{shelleqito}).  We used the ``weak-order-one Euler''
scheme, the details of which may be found in Appendix A.  Roughly,
this means interpreting the Ito equation (\ref{shelleqito}) as a
time-difference equation.  This scheme is of order one (in the time
step $\Delta t$) for averaged quantities such as the structure
functions.  Averages are calculated as time averages, assuming
ergodicity (we checked that changes in the seed of the random
generator do not affect the results).  Integrating over a large number
of realizations is thus equivalent to integrating over many large-eddy
turnover times.

First we validated our Monte Carlo simulation by comparison with 
the results for the second and fourth-order structure functions 
obtained from the moment equations in the previous section.
The comparison can be seen
in Fig.~2. The agreement of the Monte Carlo and 
moment based calculations is comparable to the machine precision
(single precision).
We mention that, in order, to maximize the extent of the inertial 
range and avoid an inertial-diffusive range with algebraic fall-off 
of the passive scalar spectrum \cite{FW}, we assumed an exponential 
cutoff on the velocity spectrum.

In all calculations $19$ shells were used, the molecular diffusivity
was varied between $2^{-10}$ and $2^{-8}$ and the time step between
$2^{-29}$ and $2^{-25}$.  In order to give an example of the quality
of the scaling, we show in Fig.~3 the log--log plot of
the $4^{th}$, $6^{th}$ and $8^{th}$ structure function versus the
$2^{nd}$ order structure function. This is done by ESS in order to improve
the scaling.


Fig.~4 shows the scaling exponents $\zeta_p$ determined by 
least square fits using ESS up to order $p=16$ for $\xi=2/3$.
The error bars are obtained from the least square fits.
In all the calculations the inertial range included eight shells or more.
When halving the number of samples in our statistics, we found that 
the values obtained remained well within the error bars.
The graph of $\zeta_p$ appears to be linear at values of $p$ beyond
$8$, with a slope of about $0.29$.
We do not rule out that the asymptotic linear trend is an 
artifact due to insufficient statistics.
We observe that the $\zeta_p$'s obtained with our Gaussian 
white-in-time velocity are less anomalous than those reported 
in Ref.~\cite{PVJ}, where the velocity was already multifractal.


\section{Conclusions}
\label{conclusion}

We proposed and studied a shell model for a randomly advected passive
scalar.  We suggest that anomalous scaling of structure functions of
order greater than two is connected to nontrivial dependence of
inertial-range properties on both integral and diffusive scales. This
is shown, for example, in the closed equation for the fourth-order
moments $\langle \theta_m \theta_m^* \theta_l
\theta_l^*\rangle$, where inertial-range shell
correlations depend explicitly on correlations between distant shells
with $m \ll l$.  The main advantage of our model is that reliable
numerical simulations become feasible. We numerically estimated
anomalous exponents for structure functions up to order $16$ and for
various scaling exponents $\xi$ of the velocity field ($ 0.2 \le \xi
\leq 2$).  Comparing our numerical results with those obtained in a
similar passive shell model
\cite{PVJ}, but advected by a multifractal velocity field, we find
large quantitative differences.  This is due to the obvious fact that
the $\theta$-statistics are strongly correlated to the
statistics of the advecting field. Nevertheless, the possibility of
writing down closed equations for correlations of any order could help
in the understanding of intermittency in more general cases.
Unfortunately, we do not see how to implement in our model the kind of
perturbation expansion done in Refs.~\cite{GK,GKB}. The main difficulty
is the lack of long-range interactions in our model which
forbids a proper definition of an eddy diffusivity in the limit  $\xi
\rightarrow 0$.  Some long-range shell models should be introduced and
studied if one wants to follow this path.

Another interesting question is to find the simplest passive-scalar
shell model which has inertial-range intermittency. For example, it is
easy to extract from (\ref{shellmodel}) an even simpler shell model,
by retaining only terms with $\theta_{m-2}^* (t) u_{m-1}(t)$ and
$\theta_{m+2}^* (t) u_{m+1}(t)$ in the advecting part.  This model has
the same properties as listed in the Introduction and some preliminary
numerical results indicate that it has similar intermittency
corrections.

Let us, finally, remark that, up to now, all the existing work on
anomalies driven by stochastic velocity field was rooted in some kind
of analysis performed in the {\it physical space}. In contrast, our
shell model exists only  in a kind of {\it Fourier space}. 
Understanding the anomalous scaling in our shell model at a
phenomenological level could help in devising a phenomenology of
intermittency. For example, ideas connected to the popular
inertial-range cascade picture could be usefully  revisited.

\vskip 0.5cm

{\bf Acknowledgements} We would like to express our gratitude to M.
Vergassola for many stimulating and interesting discussions.  We have
also benefited from extensive discussions with R. Benzi, U.~Frisch and
A.~Noullez.  This work was supported by the French Minist\`ere de la
Recherche et de la Technologie, by the European Union (Human Capital
and Mobility ERBCHRXCT920001) and by the GDR M\'ecanique des Fluides
Num\'erique.  L.B. would like to acknowledge partial support by
Minist\`ere de l'Enseignement Superieur et de la Recherche (France)
and the Observatoire de la C\^ote d'Azur where this work was
completed.


\newpage
\section*{Appendices}
\appendix
\renewcommand{\theequation}{\thesection.\arabic{equation}}
\section{The Euler scheme}
\setcounter{equation}{0}
\label{appendixa}

The weak-order-one Euler scheme \cite{KP} associated to the 
passive scalar shell model in its Ito form (\ref{shellmodel}) reads\,:
\begin{eqnarray}
 \theta_m^{n+1}& = & 
 \{1+((a_{m}b_{m+1}+b_{m}c_{m-2}) D_{m-1}\Delta t +(-a_{m}c_{m-1}
+b_{m+2}c_{m}) 
 D_{m+1}\Delta t
\nonumber \\
& +  &a_{m-1}b_{m} D_{m-2}\Delta t-a_{m+1}c_{m} D_{m+2}\Delta t- 
\kappa k_m^2\Delta t ) \} 
\theta_m^{n}
\nonumber \\
& +  &i [ a_{m}
(\theta_{m+1}^{n*}  \alpha_{m-1} \Delta W^{n*}_{m-1}
-\theta_{m-1}^{n*}  \alpha_{m+1} \Delta W^{n*}_{m+1})
\nonumber \\
& +  &b_{m}
(\theta_{m-1}^{n*}  \alpha_{m-2} \Delta W^{n*}_{m-2}
+\theta_{m-2}^{n*} \alpha_{m-1} \Delta W^n_{m-1})
\nonumber \\
& +  &c_{m}
(\theta_{m+2}^{n*} \alpha_{m+1} \Delta W^n_{m+1}
+\theta_{m+1}^{n*}  \alpha_{m+2} \Delta W^{n*}_{m+2})]
\nonumber \\
& +  & f \delta_{m,1}\Delta W^n_{f}.
\label{eulershell}
\end{eqnarray}
Here,
\begin{eqnarray}
\Delta W^n_{m}& = &\sqrt{\Delta t} \, \eta^n_{m},
\end{eqnarray}
where the  $\eta^n_{m}$ are independent identically distributed complex
random variables of the form $a+ib$ where $a$ and $b$ are independent
Bernoulli variables with values $\pm 1/\sqrt{2}$.
The Bernoulli variables are numerically generated by
a linear feed-back shift register random number generator 
(see, e.g., \cite{Knuth}).

The choice of Bernoulli variables rather than Gausssian 
variables is particularly convenient for numerical purposes.
It ensures that averaged quantities such as moments are
correct to first order in $\Delta t$, a choice
consistent with the scheme.

\section{Moment equations}
\setcounter{equation}{0}
\label{appendixode}
We give hereafter the detailed form of the equation for 
the second order moments $E_m \equiv \langle \theta_m \theta_m^* \rangle$
and the fourth-order moment 
$P_{lm} \equiv \langle \theta_l \theta_l^*\theta_m \theta_m^* \rangle$.

\begin{eqnarray}
\dot{E}_{m}
& =  & +(-2 \kappa k^2_m  
+A_{m,m}) E_{m}+A_{m,m-2} E_{m-2}+A_{m,m-1} E_{m-1}
\nonumber\\
& +  &A_{m,m+1} E_{m+1}+A_{m,m+2} E_{m+2}+F_m.\\
\label{energy}\nonumber\\
\dot{P}_{l,m}
& =  &[1-( 2 \kappa (k^2_m +k^2_l) -A_{m,m}-A_{l,l}+\delta_{l-1,m} 2 A_{m+1,m}
+\delta_{l-2,m} 2 A_{m+2,m})] P_{m,l}
\nonumber \\
& +  & A_{m,m-2} P_{l,m-2}+A_{m,m-1} P_{l,m-1}
+A_{m,m+1} P_{l,m+1}+A_{m,m+2} P_{l,m+2}
\nonumber \\
& +  & A_{l,l-2} P_{m,l-2}+A_{l,l-1} P_{m,l-1}
+A_{l,l+1} P_{m,l+1}+A_{l,l+2} P_{m,l+2}^n
\nonumber \\
& +  & \delta_{l,m} 2 (A_{m,m-2} P_{m,m-2}+A_{m,m-1} P_{m,m-1}
+A_{m,m+1} P_{m,m+1}+A_{m,m+2} P_{m,m+2})
\nonumber \\
& +  &  (\delta_{1,m} E_{m}+\delta_{1,l} E_{l}
+\delta_{1,m} \delta_{1,l} E_{1}) F_1.
\label{str4}
\end{eqnarray}
The following notation has been used.
\begin{eqnarray}
F_{1}& =  & \frac{1}{2},
\\
F_l& =  & 0  \,\,\,\,\,\,\,\,\,\, \forall  \,\,\,\,\,\,\, l \neq  1,
\\
A_{m,m-2}& =  &2 b_{m}^2  D_{m-1},
\\
A_{m,m-1}& =  & 2 (a_{m}^2  D_{m+1}+b_{m}^2 D_{m-2}),
\\
A_{m,m+1}& =  &2 (a_{m}^2 D_{m-1} +c_{m}^2 D_{m+2}),
\\
A_{m,m+2}& =  &2 c_{m}^2 D_{m+1}, 
\\
A_{m,m}& =  & -A_{m+2,m}-A_{m+1,m}-A_{m-1,m}-A_{m-2,m}.
\label{coservenergy}
\end{eqnarray}



\newpage
\begin{thebibliography}{99}
\bibitem{SREENE}K.R. Sreenivasan,
 {\it Proc. Roy. Soc. Lond.} {\bf A434}, 165 (1991).

\bibitem{UF}
U. Frisch, {\em Turbulence} Cambridge University Press (1995).

\bibitem{OC} A.M. Obukhov, {\it Izv. Akad. SSSR, Serv. Geogr. Geofiz.}
{\bf 13}, 58 (1949); S. Corrsin, {\it J. Appl. Phys. }
{\bf 22}, 469 (1951).

\bibitem{K94}
R.H. Kraichnan, Phys. Rev. Lett. {\bf 72}, 1016 (1994).

\bibitem{K68}
R.H. Kraichnan,  {\it Phys. of Fluids}  {\bf 11},  945 (1968).

\bibitem{GK} K. Gawedzki and A. Kupiainen,
 {\it Phys. Rev. Lett.} {\bf 75}, 3834 (1995).

\bibitem{KYC} R.H. Kraichnan, V. Yakhot and S. Chen, 
{\it Phys. Rev. Lett.} {\bf 75}, 240 (1995).

\bibitem{CFL}M Chertkov, G. Falkovich, 
I. Kolokolov and V. Lebedev, {\it Phys. Rev. E}{\bf 52},
4924 (1995); M. Chertkov and G. Falkovich, ``Anomalous scaling exponents
for a white-advected passive scalar'', 
submitted to {\it Phys. Rev. Lett.} (1996).

\bibitem{CFL2}M. Chertkov, G. Falkovich and V. Lebedev, ``Non-universality
of the scaling exponents of a passive scalar advected by a random flow'',
submitted to {\it Phys. Rev. Lett.} (1996).

\bibitem{GKB} D. Bernard,  K. Gawedzki and A. Kupiainen, 
``Anomalous scaling in the N-point
functions of passive scalar'', preprint (1996) chao-dyn 9601018.

\bibitem{P} A.L. Fairhall, O. Gat, V. L'vov and I. Procaccia,
``Anomalous scaling in a model of passive scalar advection: exact results''
{\it Phys. Rev. E} in press (1996).

\bibitem{E}G. L. Eyink,
 ``Intermittency and anomalous scaling of passive scalars
in any space dimension'', preprint (1996).

\bibitem{massimo} M. Vergassola, ``Anomalous scaling for passively
advected magnetic field'', {\it Phys. Rev. E} in press (1996).

\bibitem{kada} L. Kadanoff, {\it Physics Today} {\bf 48}, 11 (1995).

\bibitem{G}E.B. Gledzer, {\it Sov. Phys. Dokl.} {\bf 18},
216 (1973).

\bibitem{YO} M. Yamada  and K. Ohkitani, {\it J. Phys. Soc. Jpn.}
{\bf 56}, 4210 (1987); {\it Prog. Theor. Phys.} {\bf 79}, 1265 (1988).

\bibitem{PVJ2} M.H. Jensen, G. Paladin and A. Vulpiani, 
{\it Phys. Rev. A}{\bf 43}, 798 (1991); D. Pisarenko, L. Biferale, 
D. Courvoisier, U. Frisch and M. Vergassola, 
{\it Phys. Fluids A}{\bf 5}, 2533 (1993).

\bibitem{BBP}R. Benzi, L. Biferale and G. Parisi, 
{\it Physica D}{\bf 65}, 163 (1993).

\bibitem{KLWB}L. Kadanoff, D. Lohse, J. Wang and R. Benzi,
{\it Phys. Fluids} {\bf 7}, 617 (1995).

\bibitem{BBKT} R. Benzi, L. Biferale, R. Kerr and E. Trovatore,
 {\it Phys. Rev. E} in press (1996).

\bibitem{PVJ} M.H. Jensen, G. Paladin and A. Vulpiani, 
{\it Phys. Rev. A}{\bf 45}, 7214 (1992).

\bibitem{KP}
P. E. Kloeden and E. Platen,
{\em Numerical Solution of Stochastic Differential Equations},
Springer Verlag (1992).

\bibitem{GL}
J. G. Gaines and T. J. Lyons,
{\it SIAM J. Appl. Math} {\bf 54}, 1132 (1994).

\bibitem{Benzi}
R. Benzi, S. Ciliberto, R. Tripiccione, C. Baudet, F. Massaioli and S. Succi,
{\it Phys. Rev. E}{\bf 48}, R29 (1993).

\bibitem{FW}
U. Frisch and A. Wirth, ``Batchelor inertial-diffusive range for 
a passive scalar advected by a white-in-time 
velocity field\,: spectrum and pdf'', preprint (1996).

\bibitem{Knuth}
D. E. Knuth, {\em Seminumerical Algorithms},
Addison Wesly (1981)


\end{thebibliography}
\end{document}